\documentclass{elsarticle}

\usepackage{hyperref}

\begin{document}

\begin{frontmatter}

\title{Precision long-term measurements of beta-decay-rate ratios in a controlled environment}
%\tnotetext[mytitlenote]{Raw data can be downloaded for analysis from \href{http://www.physics.byu.edu/}{http://www.physics.byu.edu}.}

\author{S. D. Bergeson\fnref{mailto:scott_bergeson@byu.edu}}
\author{J. Peatross}
\author{M. Ware}
\address{Department of Physics and Astronomy, Brigham Young University, Provo, UT 84602}

\begin{abstract}
We report on measurements of relative beta-decay rates of Na-22, Cl-36, Co-60, Sr-90, Cs-137 monitored for more than one year. The radioactive samples are mounted in an automated sample changer that sequentially positions the five samples in turn, with high spatial precision, in front of each of four Geiger-M\"uller tubes. The sample wheel, detectors, and associated electronics are housed inside a sealed chamber held at constant absolute pressure, humidity, and temperature to isolate the experiment from environmental variations. The statistical uncertainty in the count rate approaches a few times 0.01\% with two weeks of averaging.  Other sources of error are on a similar scale.  The data are analyzed in variety of ways, comparing count rates of the various samples on one or more detectors, and comparing count rates of a particular sample across multiple detectors.  We observe no statistically significant variations in the ratios of decay rates, either annual or at higher-frequency, at a level above 0.01\%.
\end{abstract}


\end{frontmatter}

%\linenumbers

\section{Introduction}

Several studies over the last decade have addressed the possibility of annual variations in nuclear decay rates. This was first noticed in detector calibration data from the Physicalisch Technische Bundesantalt (PTB) \cite{Fischbach2009} and from Brookhaven National Laboratory (BNL) \cite{ALBURGER1986168}. After systematic influences were accounted for, the radioactive decay rates showed regular variability at the few times $10^{-4}$ level (see Table 1 in Ref. \cite{0004-637X-737-2-65}). This analysis inspired many follow-up studies, some of which observe annual and more rapid variations
\cite{Alexeyev2016,Sturrock20168, Mohsinally2016, 0004-637X-794-1-42, Sturrock201447, O’Keefe2013, Sturrock201362, 1210.3334, Sturrock201218, Sturrock2012755, Jenkins201281, 0004-637X-737-2-65, Jenkins2010332, Sturrock2010121, JavorsekII2010173, Semkow2009415, Jenkins200942, Jenkins2009407}, and some of which do not
\cite{Nahle20158, Schrader2016202, Pomme2016281, Kossert201518, Kossert201433, Meier201463, 1208.4357, Norman2009135, Cooper2009267}. It has been suggested that decay rates are influenced by both the proximity and activity of the Sun. Correlations with higher-frequency internal solar dynamics has also been identified \cite{Sturrock20168}.

Time-dependent variations in the decay rate, if existent, would likely require an explanation involving new physics outside the standard model. Variations would call into question the validity of the exponential decay of radioactive nuclei, potentially require modifications to radiation standards, and have important implications for geochronology and astrochronology \cite{Pomme2016281}. There could also be important applications. If the variations are related to the Solar neutrino flux, for example, it might be possible to use the variations as a neutrino detector, or perhaps to measure or predict solar flares \cite{Mohsinally2016,Jenkins2009407}.

Recent studies that discount the likelihood of a solar influence on decay rates have have offered the following arguments: 1) Seasonal environmental variations can influence the performance of radiation detectors \cite{Siegert19981397, Schrader2016202}. Depending on the detector type, these variations can be as large a 0.1\% \cite{ware2015}. Decay measurements in Am and Eu, for example, show that the seasonal variations in these two elements are highly correlated, but detector-specific. Some detectors show more seasonal variation than others \cite{Pomme2016281}. This indicates the importance of understanding and controlling detector errors, something that is especially important at the sub-percent level. 2) A survey of 67 decay-rate data sets, covering 24 isotopes decaying by alpha, beta-minus, beta-plus, or electron capture shows that most isotopes have at least one data set for which the seasonal variation in the activity rate is less than 0.01\% \cite{Pomme2016281}. The remaining 43 data sets showed variations above this level. The discrepancies in the data sets are consistent with argument 1) above.

In this paper, we present an analysis of newly measured decay-rate ratios. The apparatus was described recently in Ref. \cite{ware2015}. The setup is designed specifically to remove seasonal influences in detector sensitivity by tightly controlling absolute pressure and temperature of the detector environment.  Moreover, the sample-changing system allows us to divide out remaining detector-based biases by taking ratios of count-rate measurements. From our measurements of five radioactive samples, we construct 10 unique decay-rate ratios. We are unable to detect statistically significant oscillations in the decay rate ratios at any frequency with a period of a year or less above the 0.01\% level. This data adds to a growing body of evidence suggesting that solar-related variations in the nuclear activity must be below a fractional level of 0.01\%.

\section{Experimental description}

In our experiment, the radioactive samples, detectors, and electronics are housed in a sealed chamber. The pressure inside the chamber is controlled to be $700.0 \pm 0.1$ Torr, and the  wall temperature of the chamber is controlled to be $90.0 \pm 0.1~^{\circ}$F. The gas inside the chamber is N$_2$, and the humidity ranges from 3\% to 4\%.

Five different samples are placed in bismuth-lined sample holders mounted in an aluminum wheel. The wheel sequentially rotates each sample into position above four
Geiger-M\"{u}ller tube (GM) detectors once each day. The sample position relative to the detector is regulated to within 0.01 mm. The samples are Na-22, Cl-36, Co-60, Sr-90, and Cs-137. These beta-emitters were chosen because they have shown different levels of variation in previously published studies. The sample wheel  also contains an empty space so that the background signal level can be measured.

\section{Geiger-M\"{u}ller tube data}

The four GM detectors sequentially measure beta emission from samples of Na-22, Co-60, Sr-90, Cs-137, and Cl-36. These isotopes have half-lives of 2.6029, 5.2711, 28.80, 30.05, and 302000 years, respectively \cite{nuclideDotOrg}. These samples had an initial activity of nominally 1 $\mu$Ci.  We use plastic Delrin discs with different sized holes in front of the samples to limit the detector count rate for all samples to roughly 400 counts per second (cps). This count rate was chosen so that the statistical errors in the count rate would approach a level of 0.01\% in two weeks of averaging.

Each sample is positioned over each detector for four hours each day. The count data are recorded at five-minute intervals. For every sample and detector combination, we compute the average count rate over the entire four hours, and then average 14 days together for one data point.

\subsection{Deadtime correction}

The deadtime of the GM tubes is in the neighborhood 0.2 ms, depending on tube operating parameters including internal pressure, voltage, detected radiation energy. This deadtime results in a deadtime correction to the count rates of about 8\%. While this correction may seem high for measurements claiming to reach relative sensitivities at the 0.01\% level, it is nominally the same for all of the samples on each detector. Therefore in the count rate ratio, the relative importance of the deadtime correction is reduced.

Different mathematical models can be used to estimate the number of counts arriving during the avalanche recovery \cite{Lee2000731}. In our data, we correct the measured count rate $R_m$ by assuming a deadtime $\tau$ in order to find the estimated ``true'' count rate $R_t$ using the formula
\begin{equation}
  R_{t} = \frac{R_{m}}{1 -R_m \tau}.
  \label{eqn:dtcorr}
\end{equation}
As mentioned, because the count rates are similar for all of our samples, the particular details of the model are comparatively less important. We first estimate the deadtime $\tau$ using the traditional additive method. In our data analysis, we also make small corrections in the deadtimes $\tau$ so that our fitted decay rates match the known values \cite{nuclideDotOrg}. The resulting deadtimes on our four detectors are determined to be 0.250, 0.185, 0.259, and 0.181 ms. These values are consistent with those we measured using the additive method.

\subsection{Ratio measurements reduce systematic variability}

The measured count rate for each sample-detector combination depends on  the source activity, detector sensitivity and gain, discriminator levels, and quantum efficiency. The count rate is also influenced by geometric and environmental factors, such as the proximity of the detector to the sample, sensitivity to ambient pressure, electrical charging of the sample disk, and so forth. In our previous work, we showed that typical seasonal variations in ambient pressure, for example, change the GM tube count rate by typically 0.02\% per Torr, depending on both the detector and the energy of the detected beta particle. Dark signals, background levels, and deadtime also influence the measured count rate.

\begin{figure}
  \centerline{\includegraphics[width=0.9\columnwidth]{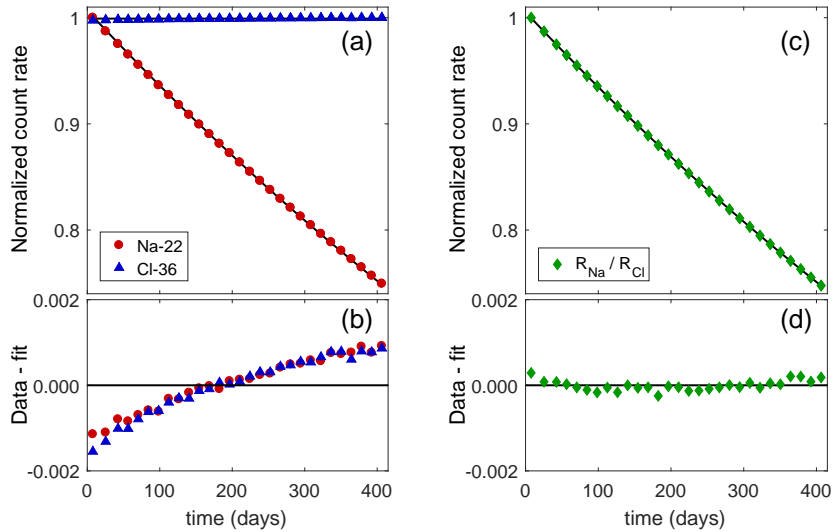}}
  \caption{\label{fig:nacl} (Color online.)
Decay rate data for Na-22 and Cl-36. (a): the count rate data for Na (red circles) and Cl (blue triangles) after dark subtraction and deadtime correction, averaged over 14 days. Also shown are exponential fits with the known lifetimes. (b): the residuals of the Na (red circles) and Cl (blue triangles) data from the upper left plot. A common-mode variation in the measured decay time compared to the known decay time is observed for both isotopes. (c): the ratio of the Na and Cl count rates (green diamonds) and the exponential fit using the  difference in known decay rates. (d): the residuals of the ratio data from plot (c). The common-mode drift in the count rate data is almost entirely removed in the ratio. The data here for Na and Cl are the averaged, normalized count rates from the four detectors.
  }
\end{figure}

A significant advantage of our experiment is that it allows us to compute count rate ratios rather than being restricted to individual sample-detector data.  In these ratios, nearly all of the factors mentioned above divide out or are significantly minimized. This is illustrated in Fig.~\ref{fig:nacl}. This plot shows relative count rate data for both Na-22 and Cl-36 as a function of time (upper left plot) after deadtime correction and background subtraction. Fitting the data using the known exponential decay rate reveals a slow time-dependent variation in the residuals (lower left plot). The data in this plot are averaged across all four detectors. The residuals for the individual detectors vary significantly, with trends that are somewhat steeper than the average shown in Fig.~\ref{fig:nacl}(b) to nearly flat. This suggests that trends in residuals are largely detector artifacts. However, when the ratio between different isotopes is taken, and the ratio data are fit to an exponential decay, this common  drift in the residuals is absent, as seen in the two right-hand plots of Fig.~\ref{fig:nacl}. This analysis clearly shows the power of the ratio technique. Systematic detector-based variations are significantly reduced. This analysis is the same as that used in Ref. \cite{ALBURGER1986168}.

\subsection{Ratio data analysis}

Using five samples, we can calculate 10 unique count rate ratios for each detector. After deadtime correction and background subtraction, we calculate our count rate ratios and then fit each ratio to the exponential decay
\begin{equation}
  s(t) = s_0 \exp(-\lambda t),
  \label{eqn:lambda}
\end{equation}
with both $s_0$ and $\lambda$ as free parameters. Our results are shown in Table \ref{tab:rates}, compared to the ratios calculated from the known decay rates for each sample.

The decay rates in Table \ref{tab:rates} are calculated using a linear, weighted least-squares fit to the log of the count rate ratio data. Each data point in the fit is the average of the ratio across the four detectors. The uncertainty in each data point for the fit is the standard deviation between the four detectors. The uncertainty indicated in the parentheses in column 3 of Table~\ref{tab:rates} is the 1$\sigma$ estimated statistical uncertainty in the fitted decay constant \cite{bevington}.

\begin{table}
\caption{\label{tab:rates} Decay rates, uncertainties, and energies for each of the ten unique ratios. The known decay rates $\lambda_0$ are the differences in the rates of the individual isotopes from Ref. \cite{nuclideDotOrg}. The fitted decay rates $\lambda$ are a least-squares fit to the averaged count rate ratios using Eq.~(\ref{eqn:lambda}). The number in parenthesis in column 3 is the 1$\sigma$ statistical uncertainty in the fitted decay rate. Additional systematic errors can be inferred from the difference between columns 2 and 3 (see also Fig. \ref{fig:ratios}). Column 4 (5) shows the principal beta energy of the isotope in the numerator (denominator) \cite{nuclideDotOrg}. Column 6 is the difference of column 4 and column 5. This data is represented graphically in Fig. \ref{fig:ratios}.}
\begin{center}
\begin{tabular}{lccccccc} \hline \hline
Ratio & $\lambda_0$ & $\lambda$ (this work) & $E_u$ & $E_{\ell}$& $\Delta E$\\
\hline
Na/Cl & 0.26630 & 0.26626(9)  & 546 & 709 & -163\\
Co/Cl & 0.13150 & 0.13208(8)  & 317 & 709 & -392\\
Cs/Cl & 0.02306 & 0.02389(8)  & 514 & 709 & -195\\
Sr/Cl & 0.02407 & 0.02390(6)  & 546 & 709 & -163\\
Na/Cs & 0.24323 & 0.24228(7)  & 546 & 514 & 32\\
Co/Cs & 0.10843 & 0.10822(9)  & 317 & 514 & -197\\
Sr/Cs & 0.00100 & 0.00009(11) & 546 & 514 & 32\\
Na/Sr & 0.24223 & 0.24238(9)  & 546 & 546 & 0\\
Co/Sr & 0.10743 & 0.10837(7)  & 317 & 546 & -229\\
Na/Co & 0.13480 & 0.13409(5)  & 546 & 317 & 229\\ \hline
\end{tabular}
\end{center}
\end{table}

\begin{figure}
\centerline{\includegraphics[width=0.6\columnwidth]{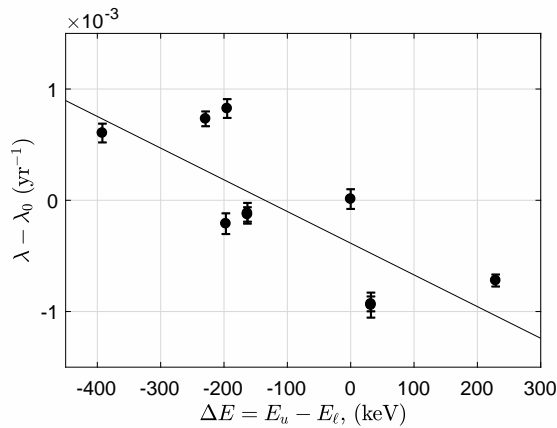}}
\caption{\label{fig:ratios} Difference of the fitted ratio decay rate, $\lambda - \lambda_0$, as a function of the energy difference $\Delta E$. The data are taken from Table \ref{tab:rates}. The error bars are the 1$\sigma$ statistical uncertainty in the fitted decay rate. The solid line is a linear fit to the data in the figure. These data suggest that the dead time correction for the GM tubes depends on the energy of the detected beta particles.}
\end{figure}

For most of the data in Table~\ref{tab:rates}, the 1$\sigma$ statistical uncertainty in the fitted decay rate ratio is smaller than the difference between the known rate and our fitted rate. However, the differences follow a systematic trend. The data from Table~\ref{tab:rates} are plotted in Fig.~\ref{fig:ratios}. The vertical axis is the difference between our fitted decay rate and the known rates, $\lambda - \lambda_0$ from Table \ref{tab:rates}. The horizontal axis is the energy difference between the beta energy from the isotope in the numerator of the ratio minus the beta energy from the isotope in the denominator if the ratio, $\Delta E \equiv E_u - E_{\ell}$ from Table \ref{tab:rates}. The data in Fig. \ref{fig:ratios} demonstrate that the exact detector deadtime is a function of the energy of the emitted particle. While this is apparent in the data, we do not attempt to correct for this effect in the data analysis.

We show this data to demonstrate the systematic uncertainties in our measurements. When measuring variations in the decay rate, it seems important to be able to find the proper nominal decay rate. The systematic variability that leads to an improper decay rate might also appear as a variability in that rate. The data in Fig. \ref{fig:ratios} suggest that our inability to properly measure the correct decay rate is on the order of $4\times10^{-4}/\mbox{yr}$, given by the standard deviation of the difference between the data and the line in Fig. \ref{fig:ratios}.

\subsection{Frequency content of the fit residuals}

\begin{figure}
\centerline{\includegraphics[width=0.9\textwidth]{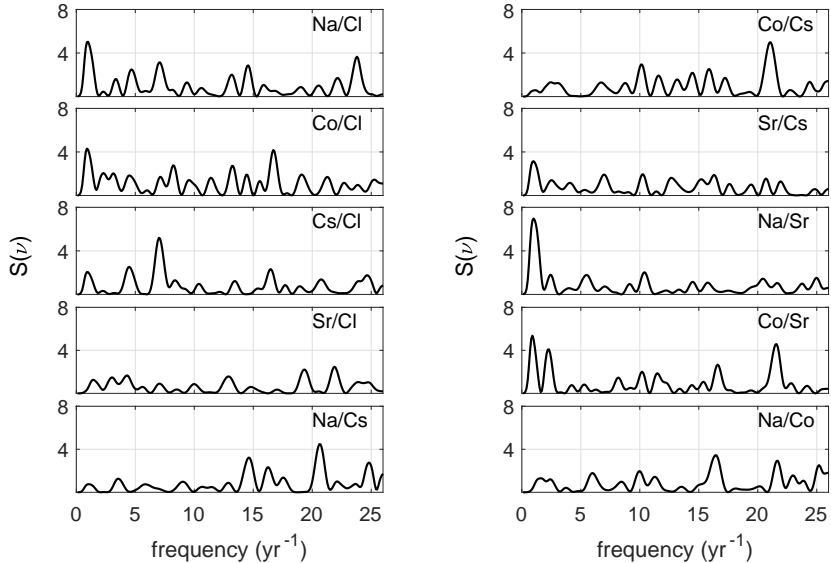}}
\caption{\label{fig:periodogram} Periodogram of the residuals of the 10 isotope ratios using Eq. (\ref{eqn:s}). The probability of a peak appearing purely randomly is given by the expression $\exp[-S(\nu)]$. Numerical simulations of random data with an rms width of $\sigma = 1.4\times 10^{-4}$ shows a random distribution of similar-sized peaks.}
\end{figure}

To look for potential time-variability in the decay rates, we average the ratio data from the four detectors, fit this data to an exponential decay, and compute the residuals of the decay ratio data to form a time series as in Fig.~\ref{fig:nacl}(d). Following the treatment in Refs. \cite{1982ApJ...263..835S, Sturrock2016}, we calculate the normalized periodogram of the residuals using the equation
\begin{equation}
  S(\nu) = \frac{1}{2\sigma^2}
    \left\{
      \frac{\left[ \sum_j r_j \cos 2\pi\nu(t_j-\tau) \right]^2}{\sum_j \cos^2 2\pi(t_j - \tau)}
      +
      \frac{\left[ \sum_j r_j \sin 2\pi\nu(t_j-\tau) \right]^2}{\sum_j \sin^2 2\pi(t_j - \tau)}
    \right\},
    \label{eqn:s}
\end{equation}
\noindent where $r_j$ are the residuals of the exponential fit, the index $j$ refers to the averaged times, and $\tau$ is defined by the equation,
\begin{equation}
  \tan (4\pi\nu\tau) = \frac{\sum_j \sin 4\pi\nu t_j}{\sum_j \cos 4\pi\nu t_j}
\end{equation}
In this equation, we use $\sigma = 1.4\times 10^{-4}$, which is approximately equal to the stability of our detectors \cite{ware2015}. A larger value of $\sigma$ reduces the size of the peaks in Fig.~\ref{fig:periodogram} quadratically. The periodograms of our 10 fitted ratios are plotted in Fig. \ref{fig:periodogram}. For comparison, we compute the periodogram of Gaussian pseudo-random noise with an rms amplitude equal to $1.4\times 10^{-4}$. This produces peaks at random frequencies but with heights that are similar to those shown in Fig.~\ref{fig:periodogram}. The traditional FFT amplitude spectra of the 10 ratios show no peaks above the 0.01\% level.  Note that for this analysis, the count rate data is averaged over 7 days instead of the 14 days to allow for analysis of higher frequencies.

\begin{figure}
  \centerline{\includegraphics[width=0.6\columnwidth]{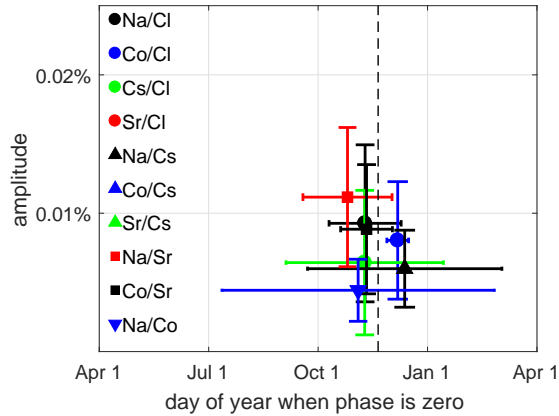}}
  \caption{\label{fig:amph} (Color online.) A plot of the amplitude, $A$, and phase, $\phi$, of the fit residuals to the function $A \cos(2\pi t/365.2422 - \phi)$, where the time $t$ is measured in days and $\phi$ is the day of the year. The amplitude error bars indicate the quadrature sum of the 1$\sigma$ standard deviation of the fitted amplitude from the four detectors together and an estimated systematic uncertainty of $1\times 10^{-4}$. The horizontal error bars show the 1$\sigma$ standard deviation of the fitted phase from the four detectors. The vertical black line corresponds November 21, the beginning of our 365-day analysis window.
  }
\end{figure}

Another way to visualize potential annual variations in the data is to fit the residuals to an annual oscillation,
\begin{equation}
  A \cos(2\pi t/365.2422 - \phi),
  \label{eqn:cosfit}
\end{equation}
with the amplitude $A$ and phase $\phi$ as free fit parameters and the time $t$ measured in days. The results are shown in Fig. \ref{fig:amph}. In the fit process, we constrained the amplitude to be positive, meaning that the relevant phase runs from $0$ to $2\pi$. The zero of phase corresponds to 12:00 am, January 1. We perform this analysis for each isotope ratio separately for each detector. The data in Fig.~\ref{fig:amph} show the average and standard deviations in $A$ and $\phi$. This analysis again indicates that none of the residuals contain a statistically significant annual oscillation above 0.01\%. The clustering of phase near November 21 corresponds to the beginning of our data analysis window which starts November 21, 2015 for this plot. This phase clustering near our start date suggests that the amplitude and phase fits are influenced by systematic errors. While we also have a few months of earlier data, those data shows somewhat larger common-mode drifts that are not canceled completely in the ratio. The analysis shown in Fig. \ref{fig:amph} includes only the most recent 365 days of data.

\section{Detrending individual detector data}

It could be argued that the ratio data automatically eliminates any possibility of detecting a solar influence in the beta-decay rate because the samples could all show the same sensitivity. However, this would also require all of the isotopes to show the same phase relative to the solar oscillation and to show a similar amplitude. On the other hand, the ratio analysis may increase the annual oscillation amplitude. In the BNL data of Ref. \cite{ALBURGER1986168}, the ratio of Si-32 to Cl-36 counts was used to determine the half-life of Si-32. The residuals in that fit showed oscillations at the 0.1\% level. A deeper analysis that incorporated both amplitude and phase fits to the individual isotope data showed that the annual oscillation amplitude was in the 0.03 to 0.05\% range. The difference in phase of the oscillation made the amplitude of the ratio larger.

One way to analyze our data without using ratios is to detrend the fit residuals in order to remove detector drift. Clearly, it is important that this detrending be accomplished without compromising a real annual oscillation. For this analysis, the count rate data of a single isotope measured on one detector are fit to an exponential decay using the known decay rates [see Fig.~\ref{fig:nacl}(a)]. The residuals to this fit, as seen, for example, in Fig.~\ref{fig:nacl}(b), are likely dominated by detector instability. These residuals are detrended using a formula of the form,
\begin{equation}
  r_2(t) = r_1(t) - [a\exp(-\gamma t) + bt + c],
  \label{eqn:r2}
\end{equation}
where $r_1(t)$ are the residuals of the proper exponential fit, and $a, b, c,$ and $\gamma$ are fit constants. The detrended residuals are then fit to an annual oscillation using Eq.~(\ref{eqn:cosfit}).

We numerically simulate the effect that such a detrending might have on an annual oscillation in the data. For the simulation, we construct 411 days of decay ``data'' with a 3.8 year half-life, normalized to unity at time $t=0$. We add four artifacts to this simulated data. First, we add random noise with a Gaussian width of 0.01\% to simulate our observed variability. Second, we add an annual oscillation with an amplitude ranging from 0.006\% to 0.1\% and random phase. Third, we add an additional decaying exponential with an amplitude of 0.1\% and a 9.5 year half-life. Finally, we add an offset of 0.1\%. We detrend this simulated data using the procedure described above. Except for a modest reduction in the fitted amplitude, we find that cosine fit reliably extracts the known amplitude and phase for the tested range.

\begin{figure}
  \centerline{\includegraphics[width=0.6\columnwidth]{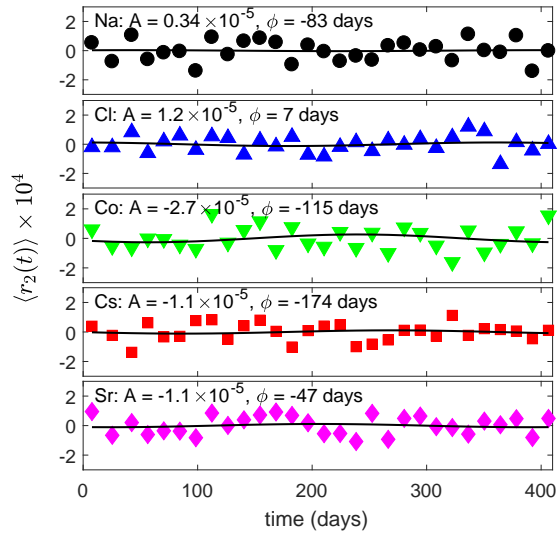}}
  \caption{\label{fig:dt} Detrended fit residuals, $r_2(t)$ from Eq.~(\ref{eqn:r2}). An isotope's detrended residuals from each of the four detectors are averaged together and then fit to an annual oscillation using Eq.~(\ref{eqn:cosfit}). The fitted amplitudes and phases relative to $t=0$ on the plot are shown.}
\end{figure}

An annual oscillation for a given isotope should be independent of the detector when the detector drift is appropriately removed. We therefore average the detrended residuals from each of the four detectors and fit this averaged detrended data to an annual oscillation. The results of this analysis are shown in Fig.~\ref{fig:dt}. The average fractional fit amplitude is $1.3 \times 10^{-5}$. Considering these five fit amplitudes as a statistical sample, and given our statistical model for fitted amplitudes of this size, this suggests an upper limit of the amplitude of the annual oscillations for any of these isotopes of approximately 0.006\%.

\section{Conclusion}

We report activity level measurements of five beta emitters using GM tube detectors for a period of over one year. We reduce detector sensitivity to seasonal variations by tightly controlling the measurement environment. The samples, detectors, and USB-powered controllers are placed inside the measurement chamber where the pressure is controlled with a precision of $\pm 0.1$ Torr and the temperature with a precision of $\pm 0.1~^{\circ}\mbox{F}$.

In one analysis, we compute the ratios of count rate measurements. These ratios divide out detector drift errors, minimizing  variations in detector response. The benefit of this analysis is shown in Fig. \ref{fig:nacl}. The fitted decay rates are close to the known rates, showing deviations from known values at the level of $4\times 10^{-4}/\mbox{yr}$. After fitting the count rate ratios to single exponential decays, we analyze the fit residuals. The frequency component corresponding to one year has an amplitude less than 0.01\%. Our spectral analysis places an upper limit of this same level for variations at any frequency.

In another analysis, we detrend the fit residuals to examine the decay behavior of individual isotopes. This treatment suggests an upper limit for a fractional annual oscillation amplitude of 0.006\%. However, it uses a phenomenological detrending model with limited physical significance.

New measurements of nuclear decay rates are clearly needed to reduce the uncertainties in decay rate measurements. Such measurements would eliminate the detector drift observed in  Fig.~\ref{fig:nacl}. They would also eliminate the need for a detrending model such as the one shown in Eq.~(\ref{eqn:r2}).  The 0.006\% to 0.01\% limit demonstrated in the present work is nearly the same as the limit demonstrated in the survey data of Ref. \cite{Pomme2016281}. That study considered data from five national standards labs spanning over six decades of measurements. New high sensitivity and high stability methods for measuring decay rates will be necessary to push the frontier of precision measurements in nuclear science.

\section{Acknowledgments}

We express appreciation to Jere Jenkins and Ephraim Fischbach for their guidance in setting up this experiment.

\bibliography{mybibfile}

\end{document}